\pdfoutput=1
\documentclass[aps,prb,superscriptaddress,nofootinbib,twocolumn,10pt,longbibliography]{revtex4-2}
\usepackage{amsmath,amssymb,amsfonts,graphicx,bm,color,xcolor}
\usepackage[colorlinks=true,allcolors=blue]{hyperref}
\usepackage[caption=false]{subfig}

\newcommand{\beq}{\begin{eqnarray}}
\newcommand{\eeq}{\end{eqnarray}}

\DeclareMathOperator{\SO}{SO}

\newcommand{\p}{\partial}
\renewcommand{\i}{\mathrm{i}}
\renewcommand{\d}{\mathop{}\!\mathrm{d}}

\newcommand{\calE}{\mathcal{E}}

\newcommand{\calM}{\mathcal{M}}

\newcommand{\CP}{\mathbb{C}P}
\newcommand{\bx}{\mathbf{x}}

\newcommand{\bn}{\mathbf{n}}

\newcommand{\NLSM}{NL$\sigma$M}

\begin{document}
\title{Magnetic D-brane solitons: skyrmion strings ending on a Néel wall in chiral magnets}
\date{October, 2025}
\author{Sven Bjarke Gudnason}
\affiliation{Institute of Contemporary Mathematics, School of
  Mathematics and Statistics, Henan University, Kaifeng, Henan 475004,
  P.~R.~China}
\affiliation{Department of Physics, Chemistry and Pharmacy, University of Southern Denmark,
Campusvej 55, 5230 Odense M, Denmark}
\author{Muneto Nitta}
\affiliation{Department of Physics 
$\&$ Research and Education Center for Natural Sciences, Keio University, 4-1-1 Hiyoshi,
  Yokohama, Kanagawa 223-8521, Japan}
\affiliation{International Institute for Sustainability with Knotted
  Chiral Meta Matter (WPI-SKCM$^2$), Hiroshima University, 1-3-1 Kagamiyama,
  Higashi-Hiroshima, Hiroshima 739-8531, Japan}
\begin{abstract}
Magnetic skyrmions extended to three dimensions form string-like
objects whose fundamental role remains largely unexplored. We show
that skyrmion strings can terminate on a Néel-type domain wall (DW),
realizing a magnetic analogue of a Dirichlet(D)-brane soliton. While
an isolated Néel DW tends to rotate into a Bloch DW, the Néel DW is
stabilized when a skyrmion string ends on it, by being
transformed into a trumpet. Unlike field-theory
D-branes, the Bloch-type DMI produces linear rather than logarithmic
DW bending, and the strings retain finite width far from the DW,
circumventing singular behavior. Furthermore, the repulsive
interaction between strings allows periodic multi-junction solutions,
yielding a square lattice of alternating strings and local DW
deformations as well as separating Bloch points.
These results establish magnetic skyrmion strings as
fundamental strings that can end on a D-brane. 
\end{abstract} 
\maketitle

\section{INTRODUCTION}

Superstring theory predicts the fundamental constituents not to be
point particles, but strings, which are generally found in two
varieties: Closed strings and open strings 
\cite{Polchinski:1998rq,Polchinski:1998rr}.
Excitations of the strings correspond to point particles at low energies.
The former need no extra ingredients in the theory and can
e.g.~describe the propagation of gravitons in spacetime.
The latter, however, predict and require the existence of hypersurfaces called 
Dirichlet(D)-branes on which 
strings terminate, 
pioneered and much developed by Polchinski \cite{Polchinski:1995mt,Polchinski:1996na}. String excitations induce  
a gauge theory on D-branes, 
described by the Dirac-Born-Infeld (DBI) action.
The discovery of D-branes was a key ingredient for the 2nd revolution of string theory in the '90ies.
A field theoretical analog of D-branes 
was proposed 
in Ref.~\cite{Gauntlett:2000de} 
as a string ending on a domain wall
(DW), referred to as a {\it D-brane soliton},  in 
the O(3) nonlinear sigma model (\NLSM) 
or the $\CP^1$ model 
with a quadratic potential, known as the easy-axis anisotropy. 
In fact, the D-brane soliton coincides with the so-called 
BIon \cite{Callan:1997kz} when it is viewed from the DBI action on the DW, 
thereby justifying to be identified with a D-brane. 
D-brane solitons were further generalized to a U(1) gauge theory \cite{Shifman:2002jm} 
and the $\CP^N$ or Grassmannian model \cite{Isozumi:2004vg,Eto:2006pg};
see also Refs.~\cite{Auzzi:2005yw,Eto:2008mf,Gudnason:2014uha}.

This brings us to consider analogous systems of D-branes in more
down-to-earth systems.
The purpose of this paper is to propose 
a D-brane-string junction in 
chiral magnets with 
a Dzyaloshinskii-Moriya-Interaction (DMI) 
\cite{Dzyaloshinskii,Moriya:1960zz} 
in three dimensions. 
Chiral magnets are known to admit various topological defects and solitons 
\cite{GOBEL20211}. 
With an easy-axis potential, 
they admit 
magnetic DWs. 
These have, in particular, been a subject of intense study due to their application to magnetic memory (storage) \cite{doi:10.1126/science.1145799,KUMAR20221}. 
 The ground state is in the ferromagnetic phase 
 for small DMI
 while for large enough DMI it is in a spiral phase or 
a chiral soliton lattice (CSL)
\cite{togawa2012chiral,togawa2016symmetry,KISHINE20151,PhysRevB.97.184303,PhysRevB.65.064433,Ross:2020orc}. 
The other more well-studied topological solitons are 
magnetic skyrmions \cite{Bogdanov:1989,Bogdanov:1995,Rossler:2006,doi:10.1126/science.1166767,doi:10.1038/nature09124,Han:2010by,doi:10.1038/nphys2045}, which are a 2-dimensional variant of the
original 3-dimensional skyrmions proposed by Skyrme himself as a
simple model for nucleons \cite{Skyrme:1962vh,Manton:2022fcb}.\footnote{
Baby-skyrmions were later considered in the 1990'ies and onward \cite{Piette:1994ug},
which differ from the magnetic skyrmions in the way they are
stabilized: The baby-skyrmions have a Skyrme term like 3D skyrmions
and the magnetic skyrmions are stabilized by the DMI.
An important consequence is that the DMI breaks parity, unlike the
Skyrme term, and this implies that only skyrmions and not
anti-skyrmions exist in the chiral magnets.
}
With a certain Zeeman magnetic field,
a skyrmion lattice is the ground state 
\cite{Rossler:2006,doi:10.1126/science.1166767,doi:10.1038/nature09124,Han:2010by,doi:10.1038/nphys2045,Ross:2020hsw}.
Isolated skyrmions have been also studied  \cite{doi:10.1126/science.1240573}. 
They have been extensively studied in particular for applications to  nanotechnology such as information carriers in magnetic storage devices \cite{Nagaosa2013}. 
See Ref.~\cite{Nagaosa2013} 
for a review of magnetic skyrmions. 
The coexistence of a DW and skyrmions in two dimensions 
gives rise to 
their composite states called DW-skyrmions~\cite{
PhysRevB.99.184412,
PhysRevB.102.094402,
Kuchkin:2020bkg,
Ross:2022vsa,
Amari:2023gqv,
Amari:2023bmx,
Gudnason:2024shv,
Leask:2024dlo,
Amari:2024jxx,
Lee:2024lge,
PhysRevB.109.014404,
Nie_2025,Saji_2025,
PhysRevB.111.024402, 
PhysRevX.15.021041, 
chen2025currentinduceddynamicsblochdomainwall} 
(see also Refs.~\cite{Kim:2017lsi,
Lee:2022rxi}) 
which are experimentally observed  
\cite{Nagase:2020imn,10.1063/5.0056100,Yang2021}.
Similar DW-skyrmions (bimerons) are proposed for an easy-plane anisotropy 
\cite{chen2024magneticbimerontravelingdomain}. 
Such DW-skyrmions are expected to be useful for manupulating skyrmions circumventing 
the skyrmion Hall effect 
\cite{zang2011dynamics,Chen:2017,Jiang:2017,litzius2017skyrmion}.

Recently, much attention has been paid to topological defects 
in three spatial dimensions; 
Hopfions \cite{Sutcliffe:2018vcb,Zheng:2023} and 
monopoles \cite{Milde:2013,PhysRevB.90.174432,tanigaki2015,fujishiro2019topological}
have been studied in chiral magnets.
In addition, 
DWs and skyrmions are extended objects in three dimensions, 
which are two dimensional hypersurfaces and strings \cite{Wolf_2021}, 
respectively.
Several aspects of skyrmion strings have been studied thus far, 
such as  
a skyrmion string ending on a monopole 
(torons) \cite{Milde:2013,PhysRevB.90.174432}, 
Kelvin modes propagating along a string \cite{PhysRevB.99.140408,PhysRevB.102.220408},
braiding skyrmion strings \cite{Zheng2021}, etc. 
One of the fundamental questions may be 
how skyrmion strings differ from or are similar to 
fundamental strings in string theory. 
We should ask whether skyrmion strings can end on a D-brane or some alternative extended structure. 
One hint to attack this problem is a D-brane soliton in field theory 
\cite{Gauntlett:2000de,
Shifman:2002jm,
Isozumi:2004vg}.

In this paper, we show that indeed skyrmion strings can end on 
a Néel DW, forming a stable D-brane soliton, 
and thus can be regarded as fundamental strings.
In this regard, the U(1) modulus of the N\'eel DW is essential and is the reason why a skyrmion string can be attached to the wall; this is not present in the Bloch DW, since the DMI induces a potential on the wall that lifts the modulus (it becomes massive)\footnote{This can be seen by the co-dimension of the DW selecting a cross product of two components of the magnetization vector and of the tangent vector in the zenith direction of the 2-sphere; for the Bloch DW the two 2-vectors are orthogonal inducing a potential for the would-be modulus, while for the N\'eel DW they are parallel so the DMI vanishes.}.
While a single Néel DW is unstable to rotate 
to a Bloch DW, it can be stabilized once a string 
terminates on it (but it is transformed into a trumpet shape).
There are outstanding differences between 
the field theory D-brane soliton 
and the magnetic D-brane soliton.
In the former, 
the string gives rise to a logarithmic bending of the DW, 
while in the latter 
we find a linear bending of the DW 
due to the Bloch-type DMI.
We also have stable finite-width strings 
far from the DW, in contrast to a singular behaviour in the former. 
We also consider the possibility of having more than one string-DW
junction.
Due to the repuslive nature of the force between strings, 
we construct a solution with periodic boundary
conditions in the ($x,y$)-plane and find a square lattice of
strings pointing up- and downwards from the DW, that locally has a
linear bending around each of the strings.
We thus establish the notion that 
magnetic skyrmion strings are fundamental strings 
that can end on a D-brane.

\section{THE CHIRAL MAGNETIC MODEL}

The ferromagnetic chiral magnetic system is described by a gradient
term (Heisenberg interaction), the DMI and a potential 
(on-site interaction) that we take
here to be the easy-axis anisotropy term
\begin{equation}
\calE = \frac12\p_i\bn\cdot\p_i\bn
+ \kappa\bn\cdot\nabla\times\bn
+ \frac12(1-n_3^2), \quad i=1,2,3,
\label{eq:E}
\end{equation}
where the total energy is given by $E=\int\calE\d^3x$ and the
magnetization vector or sigma model field, $\bn=(n_1,n_2,n_3)$, is subject to the
usual sigma model constraint: $\bn\cdot\bn=1$.
The energy and length units are rescaled so as to have a canonical
kinetic term and a potential coupling constant of unity: The only
remaining coupling constant of the theory ($\kappa$) is a combination
of the
Heisenberg exchange constant $A$, the DMI coefficient $D$ and the
anisotropy coefficient $K$, given by
\beq
\kappa = \frac{D}{2\sqrt{AK}}.
\eeq
In this paper, 
we consider the ferromagnetic phase 
$\kappa\in[\kappa^{\rm crit,junc},\kappa^{\rm crit,CSL}]$ with
$\kappa^{\rm crit,junc}\approx0.3$ being the critical coupling below
which no string-DW junction exists and
$\kappa^{\rm crit,CSL}\approx0.6$ being the critical coupling above which a CSL-type phase 
is the ground state (for $\kappa>\kappa^{\rm crit}$) 
\cite{togawa2012chiral,togawa2016symmetry,KISHINE20151,PhysRevB.97.184303,PhysRevB.65.064433}.
The chiral magnetic materials Pt/Co/Ta in
Ref.~\cite{Woo2016} correspond to $\kappa\sim0.39$, whereas CoPt
corresponds to $\kappa\sim0.43$ \cite{Sampaio2013}.\footnote{
We should point out that the materials here have N\'eel-type DMI,
whereas we have chosen Bloch-type DMI in our model.
For a discussion and comments on the Bloch- versus N\'eel-type DMI,
see App.~\ref{app:bloch_vs_neel}.
}
We also concentrate on static configurations  
and do not discuss dynamics, but point out that the fixed points of our energy-minimization are also fixed points of the dynamical Landau-Lifshitz-Gilbert equation with nonvanishing Gilbert damping.

The topology of the theory can be classified into codimensions. 
First, drawing a line in the $z$-direction, the importance of the
potential comes into light: The DW 
(of codimension one)
is supported by the existence of
two generate and disconnected vacua
\beq
\pi_0(\calM_{\rm vac})=\pi_0\big(\{n_3=1\cup n_3=-1\}\big)
\simeq 
\mathbb{Z}_2.
\eeq
Taking instead a slice, say in the ($x,y$)-plane, the topology of
2-dimensional skyrmions is given by the 2nd homotopy group that
supports also \NLSM\ lumps:
\beq
\pi_2(S^2)\simeq \mathbb{Z}\ni Q_{\rm skyrmion}.
\eeq
This $\pi_2$ corresponds to the point-compactified ($x,y$)-plane,
also known as the extended complex plane, and is topologically identical
to the Riemann sphere ($S^2$).
Finally, the 3-dimensional space mapped to the sigma model target space
being a 2-sphere, gives rise to the Hopf invariant
$\pi_3(S^2)=\mathbb{Z}\ni Q_{\rm Hopfion}$,
which supports Hopfions that are knotted string-like solutions \cite{Sutcliffe:2018vcb,Zheng:2023}
-- we will not discuss those further here. 
In this paper, we construct the string-DW junction, which is
a combination of the two former types of soliton.

\section{REVIEW OF THE FIELD THEORY D-BRANE}

Let us briefly review the field theory D-brane soliton solution, which is the string-DW
junction for $\kappa:=0$, that endows the DW with a logarithmic bending.
The solution is \cite{Gauntlett:2000de,Shifman:2002jm,Isozumi:2004vg}
\beq
\eta = r e^{\i\theta\pm z},
\eeq
where we have used the Riemann sphere coordinate
\beq
\eta = \frac{n_1+\i n_2}{1+n_3},
\eeq
on the sigma model target space and cylindrical coordinates on space:
$x+\i y=r e^{\i\theta}$.
To observe the logarithmic being of the DW, we can simply analyze the
iso-curve corresponding to the field at a fixed level set, say $|\eta|=1$:
\beq
\log r = \pm z.
\eeq
This equation also reveals that far from the junction, on the string
side, the string becomes infinitesimally small in size.

\section{THE CHIRAL MAGNETIC D-BRANE}

We now turn to the case of $\kappa\neq0$ and will for concreteness
work in the ferromagnetic phase with
$\kappa\in[\kappa^{\rm crit,junc},\kappa^{\rm crit,CSL}]$.
Firstly, we have stable skyrmion strings in this case with $\kappa>0$
turned on.
This can be understood simply by comparing each slice at a fixed $z$
to the usual 2D magnetic skyrmions and see that up to a choice of
potential, these are identical.
The DWs perpendicular to a vector in the
($x,y$)-plane 
are of the Bloch-type, for which DMI contributes 
negatively to the energy, and are stable. 
In the presence of a Bloch DW,
the skyrmion string is repelled \cite{Gudnason:2024shv} and does not automatically  
flow to the DW to become a DW-skyrmion 
\cite{
PhysRevB.99.184412,
PhysRevB.102.094402,
Kuchkin:2020bkg,
Ross:2022vsa,
Amari:2023gqv,
Amari:2023bmx,
Gudnason:2024shv,
Leask:2024dlo,
Amari:2024jxx,
Lee:2024lge,
PhysRevB.109.014404,
Nie_2025,Saji_2025,
PhysRevB.111.024402, 
PhysRevX.15.021041, 
chen2025currentinduceddynamicsblochdomainwall,Nagase:2020imn,10.1063/5.0056100,Yang2021}.\footnote{In the CSL phase, 
the Bloch DWs have negative energy 
and constitute a CSL \cite{togawa2012chiral,togawa2016symmetry,KISHINE20151,PhysRevB.97.184303,PhysRevB.65.064433}, 
and 
the DW-skyrmion string, if formed, is unstable to split into a bimeron (pair)
\cite{Amari:2023bmx}.
}
On the other hand,
the DW is of Néel-type when perpendicular to the $z$-direction
and is endowed with a U(1) modulus.
The N\'eel DW is actually unstable when compared to a Bloch DW that is
perpendicular to a vector in the $(x,y)$-plane, because the Bloch DW
has a negative contribution to the DMI energy (that fixes the would-be
U(1) modulus), whereas the N\'eel DW perpendicular to the
$z$-direction has a vanishing DMI energy.
Nevertheless, we show the Néel DW is stabilized 
in the ferromagnetic phase when a skyrmion string terminates on
it, in the sense that the N\'eel DW changes form and becomes
a trumpet -- as we shall see shortly.
One essential point is that the Néel DW has an exact U(1) 
modulus on it while the Bloch DW does not -- this facilitates that the winding of the string can be connected to the DW.

First, to establish that the bending of the wall will be linear, we
cannot simply fix the field value of the solution, since we are not
aware of analytic exact solutions to the equations of motion in this
case.
However, using the idea of fixing a curve in the equation of motion,
\beq
{\rm eom}:=
\p_i^2\eta
- \frac{2(\p_i\eta)^2\bar\eta}{1+\eta\bar\eta}
-n_3\eta
\mathop-2\i\kappa\bn\cdot\nabla\eta = 0,
\label{eq:eom}
\eeq
we need to expand the equation linearly around $|\eta|=1$
(corresponding to $n_3=0$ and hence the midpoint of the DW), arriving
at:
\beq
\p_r^2\alpha + \frac1r\p_r\alpha + \p_z^2\alpha
+ \frac{\alpha}{r^2} + \alpha
+ \frac{2\kappa(1+\alpha)}{r} = 0,
\eeq
where we have set $\eta=(1+\alpha(r,z))e^{\i(\theta+\pi/2)}$ and expanded to
linear order in $\alpha$ and we have used that
\beq
\bn = \big(\cos(\theta+\pi/2),\sin(\theta+\pi/2),-\alpha\big) + \mathcal{O}(\alpha^2).
\eeq
We are not actually interested in the soliton's behavior away from
$\alpha=0$, but simply in how the $\alpha=0$ curve could bend.
A linear partial differential equation (PDE) can be solved using the
method of the product Ansatz, but that is not suitable here since we
want to find a solution to the infinitesimal movement of the DW
position corresponding to $\alpha=0$.
We therefore propose the Ansatz $\alpha=R(r)+Z(z)$, which yields
the two ordinary differential equations (ODEs) for the position
curve:
\beq
R'' + \frac1rR' + \frac{2\kappa}{r} = \mu,\qquad
Z'' = -\mu.
\eeq
Consistency in the linearized treatment requires us to set the
constant $\mu:=0$, that otherwise would give rise to a quadratic behavior,
whereas our treatment only extends to linear and logarithmic terms.
The solution for $R(r)=-Z(z)$ corresponding to $\alpha=0$ with
$\mu:=0$ reads
\beq
\kappa r + c_r\log r + c_z z = 0,
\label{eq:curve}
\eeq
where dimensional analysis\footnote{More specifically, by reinstating units into the equations, the ratio $c_z/c_r=\pm m$ with $m^2$ being the coupling of the potential term. Our dimensionless equations correspond to $m=1$. } fixes $c_r/c_z=\pm1$ when $\kappa=0$.
This reproduces the logarithmic bending of the DW when $\kappa=0$ and
predicts a linear bending of the DW when $\kappa\neq0$.

\section{NUMERICAL RESULTS}

In order to confirm our prediction, we turn to numerical computations
of the equation of motion \eqref{eq:eom} coming from varying the energy functional
\eqref{eq:E}.
We discretize a box typically of the size $511\times511\times767$
where the derivatives of the equation of motion are approximated by a
4th order 5-point stencil finite-difference scheme, with lattice
spacing $h=0.15$.
The method for finding solutions utilized here is the so-called
arrested-Newton flow, where we solve
\beq
\p_\tau^2\eta = {\rm eom},
\eeq
monitoring the static energy given by Eq.~\eqref{eq:E} at every step
of the flow in $\tau$ ($\tau$ is a fictitious time parameter).
If the static energy rises, we stop the flow by setting
$\p_\tau\eta=0$ and then restart the algorithm.
We employ Neumann boundary conditions at the top, Dirichlet boundary conditions ($n_3=-1$) at the bottom and free boundary conditions on the vertical sides of the box.

\begin{figure}[!t]
  \centering
  \includegraphics[width=\linewidth]{{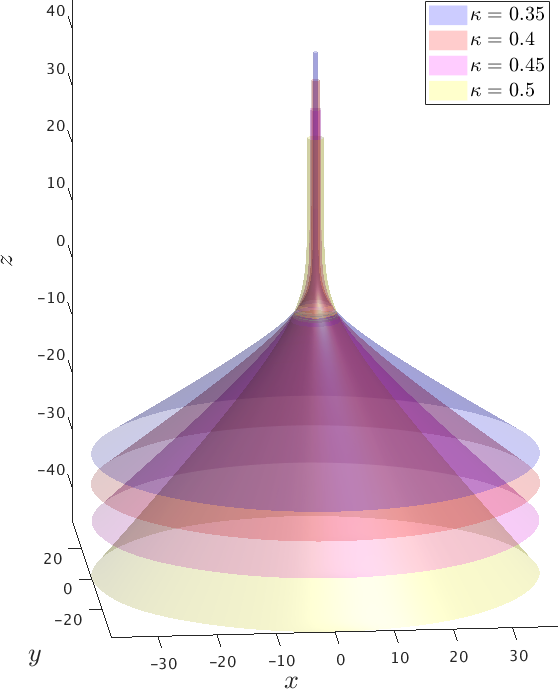}}
  \caption{String-DW junctions (trumpets) for various strength of the DMI $\kappa=0.35$ (blue), $0.4$ (red), $0.45$ (magenta),
    $0.5$ (yellow).
    The slope of the linear bending is nearly linearly proportional to
    $\kappa$.
  }
  \label{fig:stringjuncs}
\end{figure}
\begin{figure}[!t]
  \centering
  \includegraphics[width=\linewidth]{{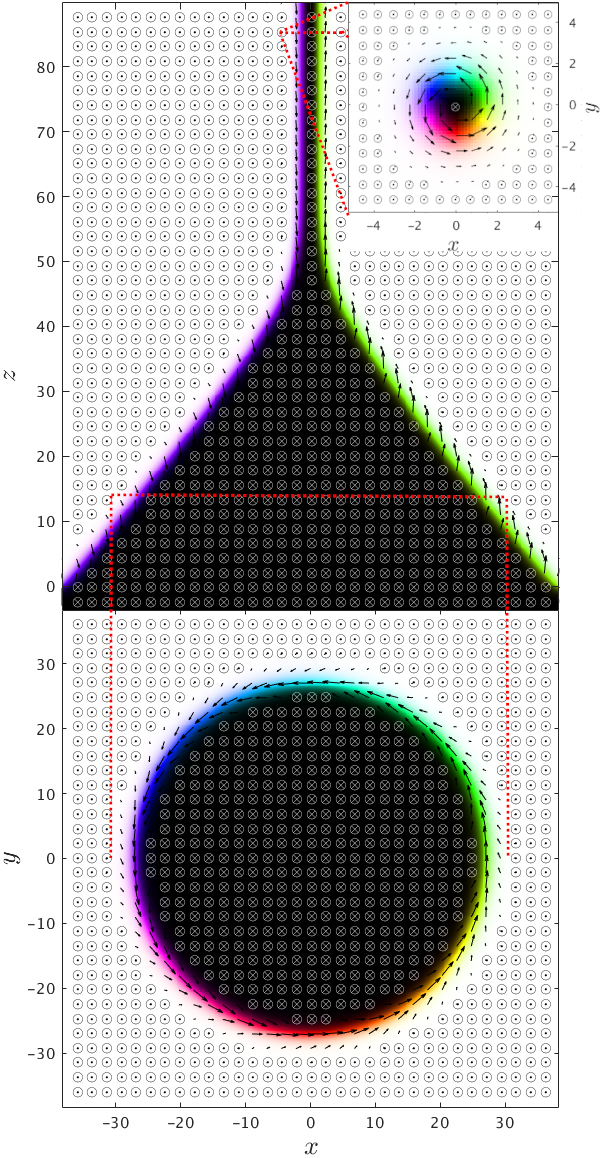}}
  \caption{String-DW junction with $\kappa=0.5$.
    Arrows are mapped from $\bn$ to $\bx$ and the colors are in
    one-to-one correspondence with the arrows: $n_1=1\mapsto$ red,
    $n_2=1\mapsto$ green, $n_1=-1\mapsto$ cyan, $n_2=-1\mapsto$
    magenta, $n_3=1\mapsto$ white and $n_3=-1\mapsto$ black.
    The inlet and the bottom are horizontal cross sections, whereas the 
    top is a vertical cross section of the string-DW junction.
  }
  \label{fig:stringjuncspin}
\end{figure}

In Fig.~\ref{fig:stringjuncs}, we display the numerical results for the
string-DW junction (trumpet) for various values of $\kappa=0.35$, $0.4$, $0.45$,
$0.5$,
illustrating the dependence of the slope of the linear bending of the
wall on the DMI coupling $\kappa$, with the steepest slopes
corresponding to the largest values of $\kappa$.
The figure shows the isosurface of the $n_3$ field at $n_3=0$; this is
the midway between the two vacua of the potential in Eq.~\eqref{eq:E}.
Clearly the DW obeys a linear relation between the $z$-coordinate and
the radius from the string junction's center when the radius is
sufficiently large (say of the order of 10 or larger).
The size (thickness) of the string also depends approximately linearly on
$\kappa$, which can be seen from Derrick's theorem (also known as a
virial law in other fields of physics) \cite{Derrick:1964ww}.
In a linear theory or an integrable theory, we would expect the
constants $c_r$ and $c_z$ to truly be constants (like in the field theory D-brane soliton case), but due to nonlinearities of
system we find they have some dependence on $\kappa$.
Fitting to the slope data of the solutions in
Fig.~\ref{fig:stringjuncs}, we find
\beq
c_z \approx 3.4 - 5.2\kappa, \quad
c_r/c_z \approx 1 + 12.6\kappa,
\eeq
which is consistent with the logarithmic bending with coefficient
$c_r/c_z=1$ when $\kappa\to0$.
\begin{figure}[!htp]
\centering
\includegraphics[width=\linewidth]{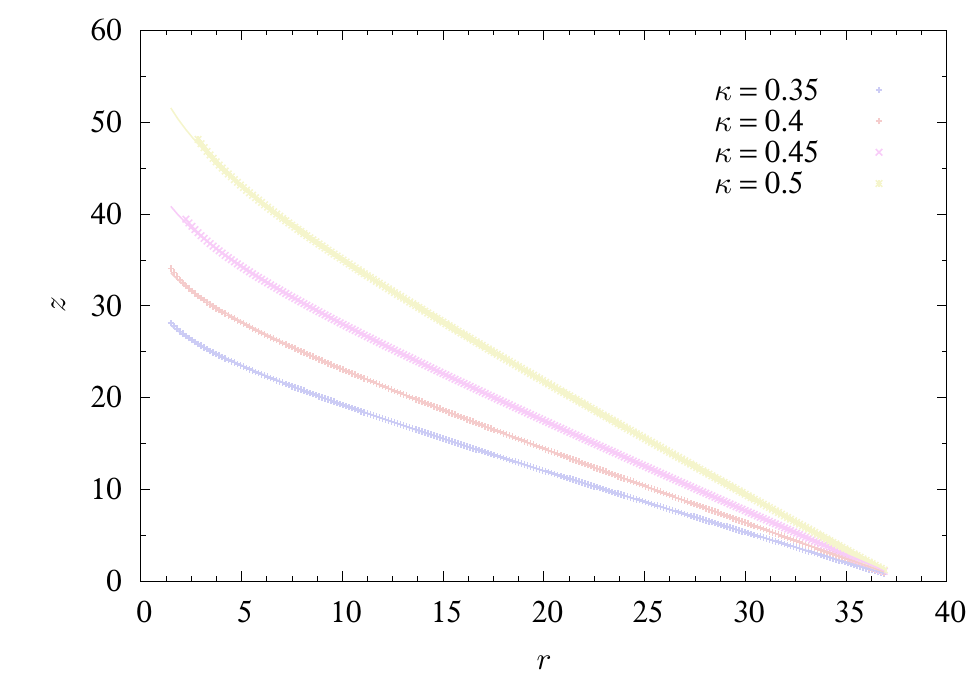}
\caption{The linear bending of the composite DW away from the string
  with $r=0$ the position of the skyrmion string. The colored points
  correspond to numerical computations of Fig.~\eqref{fig:stringjuncs}
  (colors are the same as in Fig.~\ref{fig:stringjuncs})
  whereas the curves are given by Eq.~\eqref{eq:curve} with the
  coefficients fitted to the numerical data.
}
\label{fig:curve}
\end{figure}
We show the slopes of the trumpets in an $(r,z)$-cross section in
Fig.~\ref{fig:curve}.
Interestingly, the approximate relation between $c_z$ and $\kappa$
predicts a critical value of the coupling, which indeed exists due to
the transition to the CSL: The linear fit predicts
$\kappa^{\rm crit}\approx0.65$, whereas in the full nonlinear theory
we find it to be closer to $\kappa^{\rm crit}\sim0.6$.
We find also a lower critical value for $\kappa$, where the string-DW
junction ceases to exist near $\kappa^{\rm crit,junc}\gtrsim0.3$,
although the skyrmion string itself still exists at $\kappa=0.3$.
Although the skyrmion string is stable, the DW is unstable and for
this small value of $\kappa$ it is too close to being a horizontal
N\'eel-type DW, that by the perturbation of the string collapses the
soliton junction.

Fig.~\ref{fig:stringjuncspin} shows the detailed information of the
entire 3-vector field, $\bn$, at the string-slice in the inlet at the
top-right corner, at the wall-slice at the bottom of the figure and
finally in a vertical cross section in the top of the figure.
In this figure, the vector $\bn$ is mapped onto real space
($\mathbb{R}^3$) using arrows, with the up vacuum ($n_3=1$) marked by
$\odot$, the down vacuum ($n_3=-1$) marked by $\otimes$ and the
vectors in the plane corresponding to ($n_1,n_2$) when $n_3$ is
small.
The color scheme is in one-to-one correspondence with the arrows and
is described in the caption of the figure. 
There are about 15 times more lattice points in the solution, than 
arrows depicted on the figure.

\begin{figure*}[!t]
  \centering
  \includegraphics[width=\linewidth]{{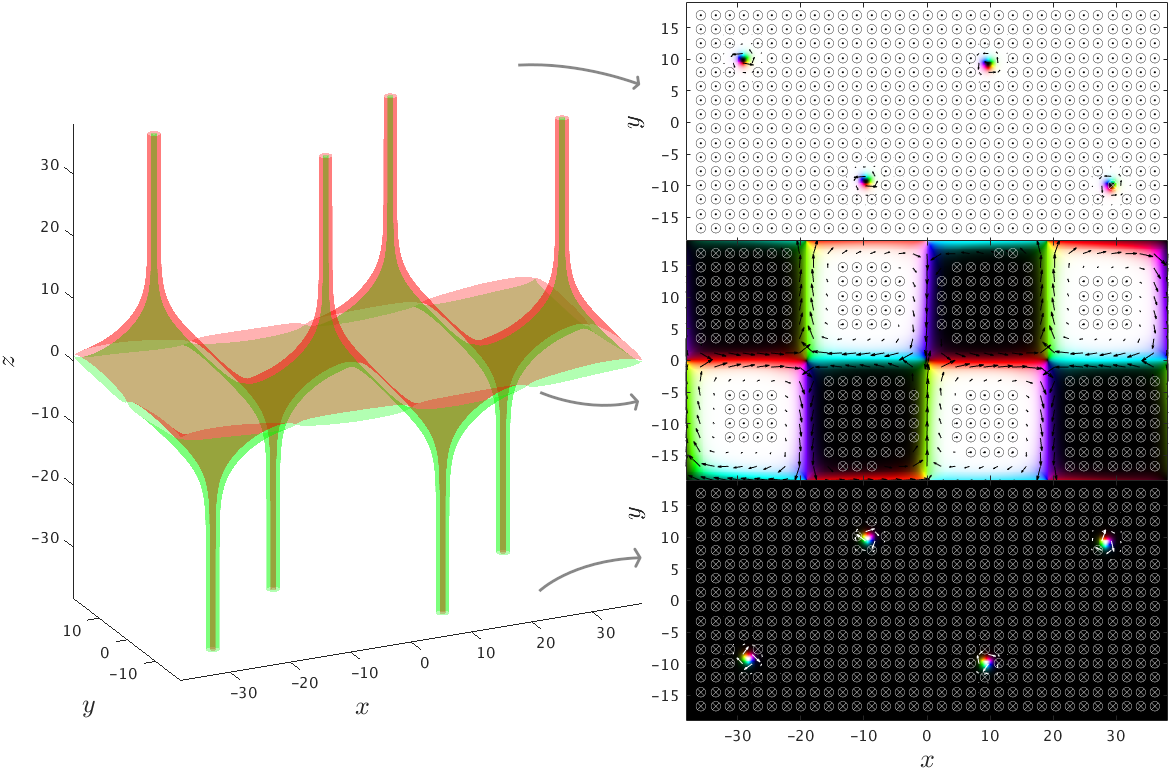}}
  \caption{Multi-string-DW junctions with $\kappa=0.4$.
    Periodic boundary conditions are imposed in the $x$- and
    $y$-directions.
    Left panel: the two isosurfaces corresponding to $n_3=1/2$
    (red) and $n_3=-1/2$ (green). Right panel: the corresponding
    details of the entire field using arrows, which are mapped from
    $\bn$ to $\bx\in\mathbb{R}^3$ and the colors corresponding to the arrows are
    described in Fig.~\ref{fig:stringjuncspin}.
    In the top-part of the configuration, the spins circle
    counter-clockwise corresponding to skyrmion strings, whereas in the
    bottom-part of they circle clockwise corresponding to
    anti-skyrmion strings.
  }
  \label{fig:stringweb}
\end{figure*}
This string junction turns out to repel other string junctions, both
of strings pointing upwards (skyrmion strings) and strings pointing
downwards (anti-skyrmion strings).
Packing the string junctions into a given volume, however, has
solutions which can most easily be found by considering periodic
boundary conditions in the ($x,y$)-plane,
although such solutions -- by topology -- naturally possess Bloch
points which locally give rise to a high-energy contribution from the DMI.
Such a configuration of multi-string-DW junctions is shown in
Fig.~\ref{fig:stringweb}, with the upper part of the space being mostly 
in the up vacuum ($n_3=1$) and the lower part mostly in the down
vacuum ($n_3=-1$).
The right-hand side of the figure shows the spin interpretation of the
map from $\bn$ to real space, see the caption of
Fig.~\ref{fig:stringjuncspin} for details of the color scheme.

\begin{figure}[!ht]
  \centering
  \includegraphics[width=\linewidth]{{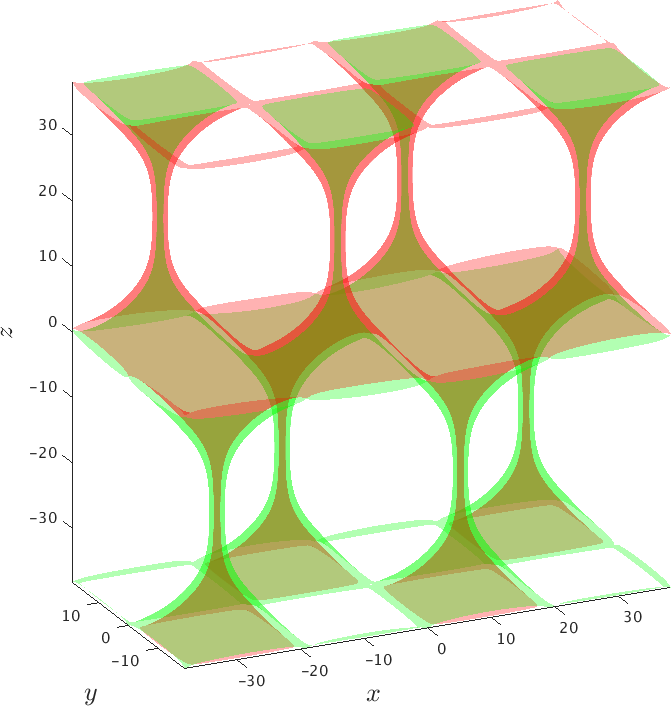}}
  \caption{
    Multi-string-DW junctions with $\kappa=0.4$, similar to
    Fig.~\ref{fig:stringweb}, but with periodic boundary conditions
    imposed in all three directions.
  }
  \label{fig:stringweb_periodic}
\end{figure}
Although we mentioned in the introduction, that the chiral DMI
supports only skyrmion strings and no anti-skyrmion string, this is
indeed the case in the up-vacuum (upper part of
Fig.~\ref{fig:stringweb}).
However, in the down-vacuum (lower part of Fig.~\ref{fig:stringweb}),
the situation is reversed and the DMI supports only anti-skyrmion
strings.

Interestingly, if we impose periodic boundary conditions also in the
$z$-direction, we obtain a fully periodic configuration just like that
of Fig.~\ref{fig:stringweb}, but with the anti-DW placed equidistantly
from the DW, see Fig.~\ref{fig:stringweb_periodic}.
This means that one could have $n$ layers of cyclic DW-anti-DW
configurations with skyrmion strings and anti-skyrmion strings
stretched between them. 

\section{DISCUSSION AND OUTLOOK}

An interesting future direction would be to study our DW-string
composite soliton at finite temperature.
It will be interesting to consider realistic boundary
conditions due to the surface physics of materials in order 
to find suitable materials as host that also must have a large
anisotropy -- finding such candidate material is essential for
future discovery of our proposed string-DW junction phase, which we
will leave for future studies.
Further interesting effects to take into account in future work, would
be to compute the backreaction from the demagnetizing field and see if
this changes the stability or the slope from linear to another form.
Looking at the any $(x,y)$-cross section, they are all unaffected by
the demagnetizing field as they are Bloch skyrmions and the trumpet is
just an enlarged Bloch skyrmion (see Ref.~\cite{Gudnason:2025gqs}
Sec.~IV, A for details).

Real-world experimentally realizable physical systems for studying
D-brane analogs etc.~are hard to come by.
It is yet unclear if and how this could be utilized for applications
in real-world spintronics.
However, the possibility of realizing a D-brane-string junction in a
chiral magnet is very interesting for fundamental physics.
One of the important directions to explore is 
how to create a D-brane soliton.
If a Néel wall moving under an applied current passes through a fixed impurity or monopole, it will be bent and eventually 
would create a D-brane soliton. 
Another direction entails  
excitations on a D-brane soliton.
Kelvin modes propagating along a skyrmion string 
\cite{PhysRevB.99.140408,PhysRevB.102.220408} 
are the so-called type-B Nambu-Goldstone modes \cite{Watanabe:2012hr,Hidaka:2012ym}
with a fractional dispersion relation 
\cite{Kobayashi:2014eqa,
Takahashi:2015caa} 
as well as 
 ripple modes on a domain wall \cite{Kobayashi:2014xua,Takahashi:2015caa,Watanabe:2014zza}.
 It will be worthwhile to study how they are transformed to each other when propagating along a D-brane soliton.
D-brane analogs have been also proposed in Bose-Einstein condensates (BECs) \cite{Kasamatsu:2010aq,Nitta:2012hy,Kasamatsu:2013lda,Kasamatsu:2013qia}, but a clear advantage of the chiral magnetic systems studied here is the stability of the magnetic system itself, as opposed to the short life-time of BECs. 
Condensed matter D-branes open up a new avenue of 
composite topological objects to simulate fundamental physics, which could be realized in other systems such as liquid crystals and optics.

\appendix

\section{Bloch versus N\'eel DMI}\label{app:bloch_vs_neel}

In our model, we have chosen Bloch-type DMI in our model, mainly not
to break $\SO(3)$ symmetry explicitly, as is done by the N\'eel-type
DMI, as that selects a special direction, namely the $z$-direction as
the spatial direction that has no derivatives in the DMI term.
This can be seen from the writing out the terms of the N\'eel DMI
\beq
\kappa\left(n_3\nabla\cdot\bn - \bn\cdot\nabla n_3\right),
\eeq
which even with $\nabla=(\p_1,\p_2,\p_3)$ contains no $z$-derivative
term due to cancellation.

It is worthwhile to point out that there is no mathematical difference
between Bloch-type and N\'eel-type DMI for axially symmetric
configurations, since the $z$-derivative in the Bloch DMI vanishes for
those.
This can be seen by employing an Ansatz
\beq
\bn = \begin{pmatrix}
  f(r,z)\cos(\theta+\delta)\\
  f(r,z)\sin(\theta+\delta)\\
  g(r,z)\sin(\theta+\delta)
\end{pmatrix},\qquad
x+\i y=r e^{\i\theta},
\eeq
for which the part of the Bloch DMI, that contains $z$ derivatives
\beq
\kappa\bn\cdot\nabla\times\bn\supset
\kappa(n_2\p_3n_1-n_1\p_3n_2) = 0,
\eeq
vanishes.

The remaining difference in the $(x,y)$-plane, between the Bloch- and
the N\'eel-type DMI, is merely a rotation by $\pi/2$, which
corresponds to a different mapping between the vectors and the
2-sphere target space.

\begin{figure}[!htp]
  \includegraphics[width=\linewidth]{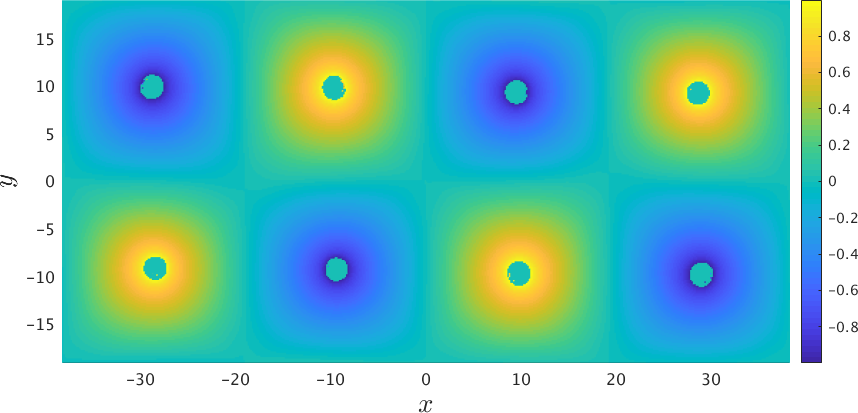}
  \caption{The absolute difference of $z$ at $n_3=0$ as function of
    ($x$, $y$) between the solution of Fig.~\ref{fig:stringweb} with
    the Bloch-type DMI and the same solution with the N\'eel-type DMI.
    The difference can be seen to be smaller than $1$ everywhere (the
    strings are vertical and therefore not present in this plot) and
    quantitatively the N\'eel-type DMI leads to about 9\% 
    steeper slopes due to the missing axial symmetry of the bent wall
    near the Bloch points.
    The linear bending is intact though, albeit with a slightly
    modified slope.
  }
  \label{fig:blochneeldiff}
\end{figure}
More quantitatively, we computed all solutions of the paper with the
N\'eel-type DMI instead of the Bloch-type DMI and confirmed that the
numerical solutions of Figs.~\ref{fig:stringjuncs},
\ref{fig:stringjuncspin} and \ref{fig:curve} with the N\'eel-type
DMI are within numerical precision of those with Bloch-type DMI --
this is due to the axial symmetry of the solutions.
However, the solutions in Figs.~\ref{fig:stringweb} and
\ref{fig:stringweb_periodic} are not axially symmetric everywhere, in which case the argument of the equivalence
(up to a remapping of the fields ($n_1\to n_2$, $n_2\to-n_1$) does
not hold.
We indeed recomputed the solutions in Figs.~\ref{fig:stringweb} and
\ref{fig:stringweb_periodic} with the 
N\'eel-type DMI and found that the slopes near the Bloch-points are
changed by about $\sim$9\%, which then continues unaltered to the
magnetic skyrmion string, due to the law of linear bending of the
N\'eel-type DWs in the presence of the string attachment, see
Fig.~\ref{fig:blochneeldiff}.

Finally, we should mention that although the Bloch skyrmion
strings are unaffected by the demagnetizing effect, the N\'eel
skyrmion corresponding to the N\'eel-type DMI feels the demagnetizing
effect and shrinks about 12\% for physically realistic values of
couplings \cite{Gudnason:2025gqs}.
This is the effect known to affect axially symmetric solitons.
We will leave the complete study of the backreaction of the
demagnetization field on the string-DW junction to future work, but we
expect the soliton to remain very similar and probably shrunk by some
percentage, depending on the numerical values of the couplings.

\bigskip
\subsection*{Acknowledgments}

S.~B.~G.~thanks the Outstanding Talent Program of Henan University for
partial support.
The work of M.~N.~is supported in part by JSPS KAKENHI [Grants
  No.~JP22H01221 and  No.~JP23K22492] and the WPI program
``Sustainability with Knotted Chiral Meta Matter (WPI-SKCM$^2$)'' at 
Hiroshima University.

\bibliographystyle{apsrev4-1}
\bibliography{references}

\end{document}